\documentclass[11pt, letterpaper]{article}
\usepackage{graphicx}
\usepackage[square,comma,numbers]{natbib}
\usepackage{amsmath}
\usepackage{amssymb}
\usepackage{amsmath}
\usepackage{amsfonts}
\usepackage{mathtools}
\usepackage[super]{nth}
\usepackage{setspace}
\usepackage{enumitem} 
\usepackage{epstopdf}
\usepackage{color}
\usepackage[letterpaper,left=1in,right=1in,top=1in,bottom=1in]{geometry}
\usepackage[ruled,vlined]{algorithm2e}
\usepackage{multirow}
\usepackage{longtable}
\usepackage{rotating}
\usepackage{mathrsfs}
\usepackage{titlesec}
\usepackage{subcaption}
\usepackage{float}
\usepackage{graphicx}
\usepackage{booktabs}
\usepackage{caption}
\usepackage{bm}
\usepackage{algorithmic}
\usepackage{tikz}
\usepackage{colonequals}
\usepackage{makecell}
\usetikzlibrary{shapes,arrows,positioning,calc}

\titleformat{\paragraph}
{\normalfont\normalsize\bfseries}{\theparagraph}{1em}{}
\titlespacing*{\paragraph}
{0pt}{3.25ex plus 1ex minus .2ex}{1.5ex plus .2ex}

\title{\Large \bf ReLU Surrogates in Mixed-Integer MPC for Irrigation Scheduling}
\author{
\centerline{\normalsize Bernard T. Agyeman$^{a}$, Jinfeng Liu$^{a}$\thanks{Corresponding author: J. Liu. Tel: +1-780-492-1317. Fax: +1-780-492-2881. Email: jinfeng@ualberta.ca.}, Sirish L. Shah$^{a}$}
\vspace{2mm}\\
\centerline{\small $^{a}$Department of Chemical \& Materials Engineering, University of Alberta,}\\
\centerline{\small Edmonton, AB T6G 1H9, Canada.}}
\allowdisplaybreaks

\begin{document}
\date{}
\maketitle

\setstretch{1.5}
{}
\begin{abstract}
	Efficient water management in agriculture is important for mitigating the growing freshwater scarcity crisis. Mixed-integer Model Predictive Control (MPC) has emerged as an effective approach for addressing the complex scheduling problems in agricultural irrigation. However, the computational complexity of mixed-integer MPC still poses a significant challenge, particularly in large-scale applications. This study proposes an approach to enhance the computational efficiency of mixed-integer MPC-based irrigation schedulers by employing ReLU surrogate models to describe the soil moisture dynamics of the agricultural field. By leveraging the mixed-integer linear representation of the ReLU operator, the proposed approach transforms the mixed-integer MPC-based scheduler with a quadratic cost function into a mixed-integer quadratic program, which is the simplest class of mixed-integer nonlinear programming problems that can be efficiently solved using global optimization solvers.
	The effectiveness of this approach is demonstrated through comparative studies conducted on a large-scale agricultural field across two growing seasons, involving other machine learning surrogate models, specifically Long Short-Term Memory (LSTM) networks, and the widely used triggered irrigation scheduling method. The ReLU-based approach significantly reduces solution times---by up to 99.5\%---while achieving comparable performance to the LSTM approach in terms of water savings and Irrigation Water Use Efficiency (IWUE). Moreover, the ReLU-based approach maintains enhanced performance in terms of total prescribed irrigation and IWUE compared to the widely-used triggered irrigation scheduling method.
\end{abstract}
\clearpage
\section{Introduction}
Agriculture is the largest consumer of freshwater globally, accounting for 70\% of the total supply~\cite{unwater2015}. The growing freshwater scarcity crisis, intensified by climate change and rapid population growth, has placed significant strain on global freshwater resources. While agriculture is significantly affected by this scarcity, it also contributes to the problem through its extensive water use. To address this challenge, implementing efficient water management strategies in agricultural irrigation is essential for mitigating the water scarcity problem and ensuring the sustainability of the agricultural sector. Closed-loop irrigation scheduling, which utilizes feedback to provide precise water amounts to crops at optimal times, has emerged as a promising solution for achieving efficient water management in irrigation.

Scheduling, in general, involves the optimal allocation of finite resources over a specified time horizon to achieve a particular objective. Irrigation scheduling, typically conducted on an hourly or daily basis, is important for determining the optimal amount of water to supply to crops and the appropriate timing for irrigation. In the context of daily irrigation, which is the primary focus of this work, the objective is to identify the specific days within a given scheduling horizon on which irrigation should be performed and to determine the precise amount of water to be applied during each irrigation event. Recently, irrigation scheduling approaches have evolved to incorporate management zones (MZs) in their designs. MZs are defined as distinct areas within a large-scale field characterized by uniform soil and crop properties. By integrating MZs, irrigation scheduling schemes can better account for the significant spatial variability present in agricultural fields. When daily irrigation scheduling is required for a field with multiple MZs, the task expands to determining the specific days within the scheduling horizon for irrigation events and identifying the optimal irrigation amounts for each MZ.

By definition, addressing the daily irrigation scheduling problem involves discrete decision-making, particularly in determining which days within the scheduling horizon should be designated for irrigation. This discrete decision-making process can be represented by assigning a binary decision variable to each day within the scheduling horizon. A value of 1 indicates that irrigation should be performed on that specific day, while a value of 0 signifies that no irrigation should occur. Alongside these discrete decisions regarding the timing of irrigation, continuous decisions must also be made concerning the precise amounts of water to be applied to the various MZs on the days when irrigation is scheduled. To effectively manage both discrete and continuous decisions, and to leverage the success of model predictive control (MPC) in addressing scheduling problems across various domains~\cite{risbeck2018closed,subramanian2012state,yi2015adaptive,delgoda2016irrigation,nahar2019closed}, mixed-integer MPC has emerged as an attractive optimal control framework for solving the daily irrigation problem.

Mixed-integer MPC for irrigation scheduling, initially proposed in~\cite{agyeman2023lstm}, leverages agro-hydrological models such as the Richards's equation, water-balance models, which describe soil moisture dynamics in agricultural fields, to determine irrigation schedules. These schedules involve `yes/no' decisions for irrigation timing and continuous decisions for irrigation amounts, aiming to maintain soil water content within an optimal range to promote crop growth while minimizing both fixed and variable costs associated with the irrigation process. This optimal control framework has been evaluated under various weather conditions and irrigation management practices, demonstrating its ability to conserve irrigation water while enhancing crop yield compared to the widely used triggered irrigation scheduling approach. However, despite its promise, several near-term improvements, particularly in enhancing the computational efficiency of the mixed-integer MPC-based scheduler, are necessary to facilitate its adoption in practical irrigation settings.

Mixed-integer problems are classified as NP-hard. Consequently, despite advancements in optimization techniques and computing power, directly applying mixed-integer solvers to the mixed-integer problems arising in mixed-integer MPC-based irrigation scheduling is only feasible for small-scale scenarios. The requirement to generate irrigation schedules within a reasonable time frame, coupled with the need to extend mixed-integer MPC formulations to fields comprising several MZs, necessitates the development of  efficient solution methods.

Several approaches have been proposed to address the computational challenges associated with mixed-integer MPC-based schedulers. Notably, the initial work~\cite{agyeman2023lstm} that introduced mixed-integer MPC for daily irrigation scheduling employed two key strategies to enhance computational efficiency. First, instead of directly using a mechanistic agro-hydrological model, such as the Richards equation—--which was found to render the mixed-integer MPC framework intractable for moderate prediction horizon values—--a long short-term memory (LSTM) network representation of the agricultural field was utilized within the mixed-integer MPC framework. The use of this surrogate model significantly improved the computational efficiency of the framework. To further enhance efficiency, the study proposed an approach that transformed the mixed-integer nonlinear problem (MINLP) resulting from the MPC formulation into a nonlinear program (NLP). Specifically, the logistic sigmoid function was employed to represent the binary decision variables. This transformation from MINLP to NLP enabled the problem to be solved within a reasonable time frame while also accommodating multiple MZs in the framework. However, despite these benefits, the use of the logistic sigmoid function introduced challenges, including approximation errors and difficulties in result interpretability. Determining the appropriate slope for the sigmoid function proved to be nontrivial, and large slopes often led to ill-conditioned optimization problems.

Two recent studies~\cite{agyeman2024semi, agyeman2024learning} have leveraged the success of reinforcement learning (RL), particularly multi-agent RL (MARL), to enhance the computational efficiency of mixed-integer MPC-based irrigation schedulers. In these approaches, multiple RL agents were trained for the various MZs within the field. These agents were tasked with determining the irrigation timing across the scheduling horizon, allowing decentralized MPCs with continuous controls to be used for calculating the irrigation amounts for each MZ. These RL-based approaches effectively addressed the interpretability issues associated with the logistic sigmoid function and facilitated the use of parallel computing to solve the decentralized MPCs, thereby improving overall computational efficiency.

While these RL-based approaches have shown promise in enhancing computational efficiency and addressing interpretability issues, they come with some limitations. One notable challenge is the significant computational resources required for training multiple RL agents, particularly in fields with a large number of MZs. The complexity of the training process can lead to prolonged training times and increased computational costs. Moreover, if there is a need to redesign the reward function, which is common for irrigation scheduling since different management practices which affect the reward of the agents have to be carried out during a particular growing season, the agents must be retrained, which can be a time-consuming process. Given the downsides associated with using RL agents to address the computational challenges of mixed-integer MPC-based schedulers, exploring efficient solution approaches that do not rely on RL remains of vital importance.

As previously mentioned, the objective of the mixed-integer MPC-based irrigation scheduler—to maintain root zone soil moisture within an optimal range while minimizing fixed and variable irrigation costs—is typically modeled with a quadratic cost function. However, the nonlinear nature of the models used to describe soil moisture dynamics—such as the Richards equation or an LSTM network representation—results in a MINLP. By exploring surrogate models that can be represented with linear constraints, it is possible to reformulate the problem as a Mixed-Integer Quadratic Program (MIQP), which is a simpler and more tractable class of MINLPs.

A ReLU (Rectified Linear Unit) network is a neural network where the ReLU activation function is applied to the neurons, except for those in the output layer. These networks have been used to model complex relationships and have been found to outperform other activation functions, such as the sigmoid function, in deep neural networks~\cite{glorot2011deep}. ReLU networks have achieved the best results on various benchmark problems across different domains (e.g., \cite{he2015delving,dahl2013improving,krizhevsky2012imagenet}), including in the modeling of soil moisture dynamics~\cite{adeyemi2018dynamic,achieng2019modelling}.

ReLU networks are composed of max-affine spline operators and are piecewise-linear and continuous functions~\cite{balestriero2018mad}. This characteristic allows a ReLU network to be exactly represented by a set of mixed-integer linear (MIL) constraints. An exact MIL formulation of a ReLU network can be achieved by representing each ReLU operator with a binary variable and applying the big-M method. The MIL formulation has been employed to enhance computational efficiency in optimization problems that embed ReLU neural networks, including applications such as learning the region of attraction of nonlinear systems~\cite{chen2021learning}, chemical process optimization~\cite{stinchfield2022optimization}, and optimizing the operation of monoclonal antibody production~\cite{obiri2023optimizing}. Despite the advancements in model predictive control-based irrigation scheduling, the literature has yet to exploit the MIL formulation of ReLU neural networks to enhance the computational efficiency of these schedulers. Furthermore, existing research on mixed-integer MPC-based schedulers has not explored the potential of using the MIL formulation of ReLU networks to improve the computational efficiency of mixed-integer MPC-based irrigation schedulers.

The main contribution of this work is to enhance the computational efficiency of mixed-integer MPC-based irrigation schedulers by leveraging the MIL formulation of ReLU networks. In this approach, the model representing soil moisture dynamics within the MPC framework is replaced with a ReLU neural network. By utilizing the MIL formulation of the ReLU network, and given that the cost function of the MPC framework is quadratic, the resulting problem is a MIQP, for which global optimization solutions exist~\cite{belotti2009branching, belotti2013mixed}. This MIQP formulation is then solved to determine the daily irrigation schedules, including both the timing and the irrigation amounts for the various MZs within the field.

Additionally, a comparative study is conducted over two growing seasons on a large-scale, spatially variable field to demonstrate that the proposed approach significantly enhances the computational efficiency of mixed-integer MPC-based irrigation schedulers compared to those utilizing an LSTM network representation of the field model. The comparison with the LSTM model is justified by its established ability to improve computational efficiency over traditional models like the Richards equation, while also maintaining superior water-saving performance compared to the widely used triggered irrigation scheduling method. Furthermore, the study includes a comparison with the triggered irrigation scheduling scheme to provide a basis for adopting the mixed-integer MPC with ReLU surrogate model in practical irrigation scheduling settings, given the triggered approach's popularity and ease of implementation in practice.

\section{System Description}
\begin{figure}[!ht]
	\centering
	\includegraphics[width=1\columnwidth]{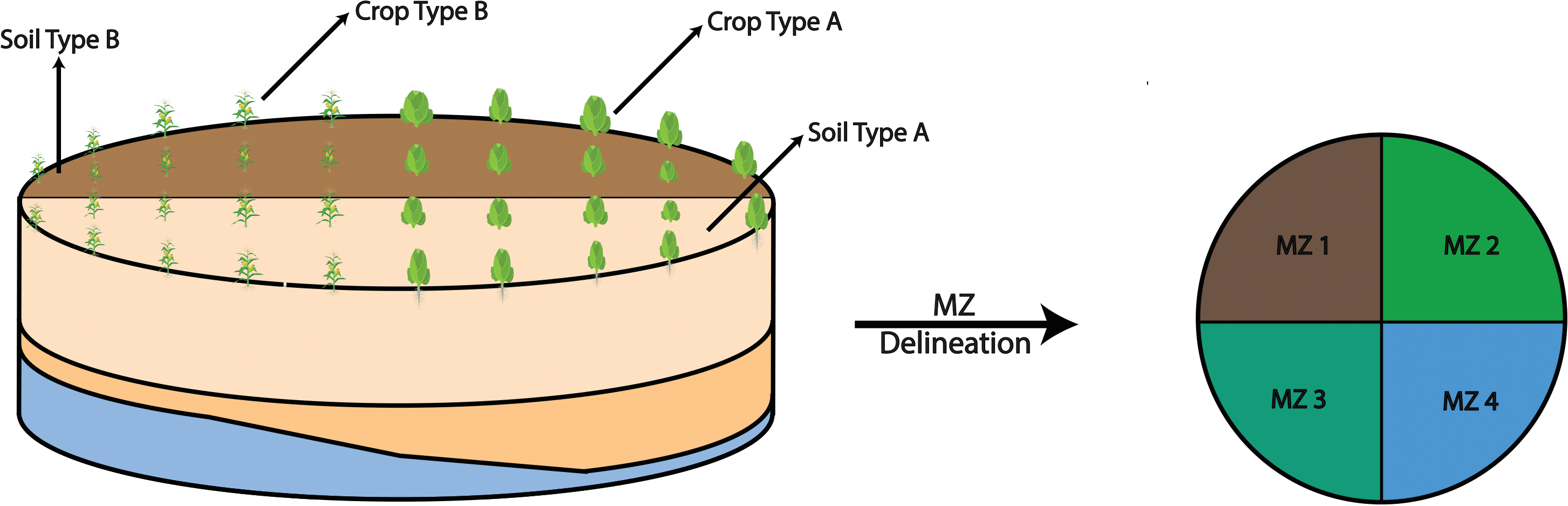}
	\caption{A schematic diagram of a spatially variable field with variability in crop and soil. The field is divided into 4 distinct MZs, each with uniform soil and crop properties.}
	\label{fig:spat_var_field}
\end{figure}
The system studied in this work is a large-scale agro-hydrological field characterized by significant spatial variability in attributes such as soil type, crop type, and elevation. To account for this spatial variability, the field is divided into distinct  MZs. Figure~\ref{fig:spat_var_field} illustrates a field with variability with respect to soil and crop. Given the presence of 2 soil types and 2 crop types, and considering that each MZ should have similar soil and crop properties, the field is delineated into 4 distinct MZs based on these attributes.

The proposed soil moisture modeling framework is particularly well-suited for a delineation approach that incorporates key attributes such as elevation and soil hydraulic parameters. Elevation influences the movement and distribution of water across agricultural fields, thereby affecting irrigation efficiency. Similarly, soil hydraulic parameters directly impact soil moisture dynamics and plant available water. A delineation approach that incorporates soil hydraulic parameters allows for the use of mechanistic models, which rely on these parameters as key inputs, to accurately model soil moisture dynamics within the various MZs. Additionally, incorporating elevation ensures that the resulting MZs have relatively flat profiles, enabling the use of mechanistic models that focus on vertical soil moisture dynamics while ignoring lateral movements.

The MZs in large-scale fields are generally known to exhibit weak interactions with each other, meaning that the impact of adjacent zones on each other's moisture dynamics is minimal. Given this phenomenon, the soil moisture dynamics can be modeled independently for each MZ in the field. 

In this study, the soil moisture dynamics in each MZ is modeled using the 1D version of the Richards equation in the axial direction. The Richards equation~\cite{richards1931capillary} is a mechanistic model that accounts for various hydrological processes, including irrigation, rain, evapotranspiration, infiltration, root-water extraction,  and surface runoff, and it has proven successful in describing soil moisture dynamics in agricultural fields~\cite{farthing2017numerical}. The 1D Richards equations used to model the soil moisture dynamics in the different MZs are calibrated with soil hydraulic parameters representative of their respective zones. For example, when clustering techniques are employed for MZ delineation, the soil hydraulic parameters for calibrating the 1D Richards equation can be those occurring at the centroid of the various clusters/MZs.

The 1D version of the Richards equation is expressed mathematically as:
\begin{equation}\label{eq:RE_1D}
	C(\psi)\frac{\partial \psi}{\partial t} = \frac{\partial}{\partial z}\bigg[K(\psi)\bigg(\frac{\partial \psi  }{\partial z}+ 1\bigg)\bigg]-\rho(\psi)\mathcal{R}\left(\text{K}_{\text{c}}, \text{ET}_0, \text{z}_{\text{r}}\right)
\end{equation}
In Equation~(\ref{eq:RE_1D}), $\psi~(m)$ is the capillary pressure head, $K(\cdot)~(m \cdot s^{-1})$ is the unsaturated hydraulic water conductivity and  $ C(\cdot)~(m^{-1})$ is the capillary capacity. 
The term $\rho(\cdot)~(-)$ is a dimensionless stress water factor, $\mathcal{R}(\cdot)$ is the root water uptake model which is a function of the crop coefficient $\text{K}_{\text{c}}~(-)$, the reference evapotranspiration $\text{ET}_0 ~(m \cdot s^{-1})$, and the rooting depth $\text{z}_{\text{r}}~(m)$. For a detailed description of the functions $K(\cdot)$ and $C(h)$, readers are referred to~\cite{van1980closed}. Similarly, for a thorough description of $\rho$ and $\mathcal{R}$, readers are referred to~\cite{feddes1982simulation}.

The following boundary conditions are imposed for the numerical solution of the 1D Richards equation:
\begin{alignat}{2}
	&\frac{\partial (\psi+z)}{\partial z}\bigg|_{z=H_z}&&=1\label{eq:botBC} \\
	&\frac{\partial \psi}{\partial z}\bigg|_{z=0}&&=-1-\frac{u^{\text{irr } } - \text{EV}}{K(\psi)}\label{eq:topBC}
\end{alignat}
where $H_z~(m)$, $u^{\text{irr}}~(m \cdot s^{-1})$, and $\text{EV}~(m\cdot s^{-1})$ in Equations (\ref{eq:botBC}) and (\ref{eq:topBC})  represent the depth of the soil column, the irrigation amount and the evaporation rate, respectively. A detailed discussion of the numerical solution of Equation~(\ref{eq:RE_1D}) is omitted. Interested readers are referred to~\cite{bo2020parameter} for an in-depth discussion on the method of lines (MOL) numerical approach, which is employed in this work to solve the 1D Richards equation.

After solving Equation~(\ref{eq:RE_1D}) to obtain $\psi$, the corresponding volumetric soil moisture content $\theta_v$ can be calculated as follows:
\begin{equation}\label{eq:thetareln}
	\theta_{{v}} (\psi)=\theta_r +(\theta_{{s}}-\theta_{{r}})\bigg[\frac{1}{1+(-\alpha \psi)^n}\bigg]^{1-\frac{1}{n}}
\end{equation}
where $\theta_s$ and $\theta_r$ denote the saturated and residual soil moisture contents, respectively. The terms $\alpha$ and $n$ are curve fitting parameters used to characterize the soil water retention curve. The saturated hydraulic conductivity $K_s$, along with $\theta_s$, $\theta_r$, $\alpha$, and $n$ are the relevant soil hydraulic parameters needed to solve the Richards equation. In this work, these 5 parameters are collectively denoted as  $\bm{\phi} = \left[ K_s, \theta_s, \theta_r, \alpha, n \right]$.

After applying the MOL numerical approach, the 1D Richards equation  can be expressed in state-space form as:
\begin{align}
	x_{k+1}&=\mathcal{F}(x_k,u_k, \bm{{\phi}}) + \omega_k \label{eq:state_equation}\\
	y_{k}&=\mathcal{H}(x_k,\bm{{\phi}}) + v_k \label{eq:output_equation}
\end{align}
where $x_k\in \mathbb{R}^{N_x}$ represents the state vector containing $N_x$ capillary pressure head values for the spatial nodes in $H_z$. In Equations~(\ref{eq:state_equation}) and~(\ref{eq:output_equation}), $\mathcal{F}$ and $\mathcal{H}$ represent the system dynamics and output function, respectively. The input vector $u_k$ contains irrigation amount, precipitation, reference evapotranspiration, the crop coefficient, and the rooting depth. The terms $\omega_k$ and $v_k$ denote the uncertainties in the state and output equations, respectively. In this work, the volumetric water content $\theta_{{v}}$ is denoted as the output $y_k$, making Equation~(\ref{eq:output_equation}) a general representation of Equation~(\ref{eq:thetareln}). Consequently, $y_k\in \mathbb{R}^{N_y}$ represents the output vector containing $N_y = N_x$ volumetric soil moisture content values for the corresponding spatial nodes in $H_z$.

Given that the soil moisture dynamics are modeled independently for the MZs,  for a field with $M$ MZs,  the state vector ${\bm{X}}_k \in \mathbb{R}^{N_x\times M} $ and the output vector ${\bm{Y}}_k \in\mathbb{R}^{N_y\times M} $  can be compactly represented as follows:

\begin{equation}\label{eq:state_delineated_field}
	{\bm{X}}_{k+1} = 
	\begin{bmatrix}
		{x}_{k+1, 1}\\
		{x}_{k+1, 2}\\
		\vdots \\
		{x}_{k+1, M}\\
		
	\end{bmatrix}  =  \begin{bmatrix}
		\mathcal{F}\left({x}_{k, 1}, u_{k,1}, \bm{\phi}_1\right) + \omega_{k,1} \\
		\mathcal{F}\left({x}_{k, 2}, u_{k,2}, \bm{\phi}_2\right)+ \omega_{k,2}\\
		\vdots \\
		\mathcal{F}\left({x}_{k, M}, u_{k,M}, \bm{\phi}_M\right) +\omega_{k,M}\\
	\end{bmatrix}
\end{equation}

\begin{equation}\label{eq:output_delineated_field}
	{\bm{Y}}_{k} = 
	\begin{bmatrix}
		{y}_{k, 1}\\
		{y}_{k, 2}\\
		\vdots \\
		{y}_{k, M}\\
		
	\end{bmatrix}  =  \begin{bmatrix}
		\mathcal{H}({x}_{k, 1}, \bm{\phi}_1) + v_{k,1} \\
		\mathcal{H}({x}_{k, 2}, \bm{\phi}_2)+ v_{k,2}\\
		\vdots \\
		\mathcal{H}({x}_{k, M},\bm{\phi}_M) + v_{k,M}\\
	\end{bmatrix} 
\end{equation} 
where $\bm{\phi}_i$ represents the set of hydraulic parameters that are representative of the soil type in in MZ $i$. Additionally, $x_{k, i}$, $u_{k,i}$ and $y_{k,i}$ represent the state, input and output vectors of MZ $i$ at time instant $k$, while $\omega_{k,i}$ and $v_{k,i}$ represent the uncertainties in the state and output equations, respectively.

\section{Problem Formulation}
This work addresses the daily irrigation scheduling problem for a large-scale agricultural field delineated into distinct irrigation MZs. The objective is to a develop  scheduling framework that determines both the irrigation timing (which is encoded with a binary variable) and the specific irrigation application amounts for each zone. The problem is formulated as follows:

\textbf{Given:}
\begin{itemize}
	\item \textbf{Scheduling Horizon}: The known number of days for which the irrigation needs to be scheduled.
	\item \textbf{Management Zones ($M$)}: The field is divided into $M$ distinct MZs.
	\item \textbf{Weather Predictions}: Daily predictions of reference evapotranspiration and precipitation, essential for determining the water needs of crops.
	\item \textbf{Crop Information}: This includes the crop coefficient, derived from empirical relations calibrated specifically for the crop and field under study. 
	\item \textbf{Soil Moisture Content}: The initial distribution of soil moisture content within the rooting depth across each MZ at the start of the scheduling horizon.
\end{itemize}

Determine:
\begin{itemize}
	\item \textbf{Irrigation Timing}: The days within the scheduling horizon on which irrigation should be performed. In line with standard irrigation practices, the timing should be uniform across all MZs of the field.
	
	\item \textbf{Irrigation Amounts}: The daily irrigation amount for each MZ on each day within the scheduling horizon. The scheduler should ensure that the irrigation amounts align with the timing—prescribing zero irrigation amounts on days when irrigation should not be performed and non-zero amounts on days when irrigation is scheduled.
\end{itemize}

\section{Mixed-Integer MPC for Daily Irrigation Scheduling}
A mixed-integer MPC with zone control (zone objectives), as originally proposed in~\cite{agyeman2023lstm}, is presented to address the daily irrigation scheduling problem. This MPC formulation uses discrete (binary) controls to encode the irrigation timing and continuous controls to determine the irrigation amounts that ought to be applied to the various MZs that make up the field. The scheduler's objective is to provide irrigation schedules that maintain the root zone soil moisture content in each MZ within an optimal range, ensuring optimal crop development. Additionally, the scheduler seeks to minimize the fixed and variable costs associated with the irrigation events.

For  scheduling horizon $N$,  the scheduler $\mathbb{P}_{N,M}(d)$ for a field with $M$ MZs is formulated for a paricular day $d$ as follows:

\begin{subequations}
	\begin{alignat}{3}
		&\min_{\bm{\bar{E}, ~\underline{E},~ U,~ c}} ~ \sum_{k=d+1}^{d+N}\sum_{j=1}^{M}\left[\bar{Q}_j\bar{\epsilon}_{k,j}^2 + \underline{Q}_j\underline{\epsilon}_{k,j}^2 \right] + &&\sum_{k=d}^{d+N-1}R_cc_{k} +  &&\sum_{k=d}^{d+N-1}\sum_{j=1}^{M}R_uu_{k,j}^{\text{irr}}\label{eq:cost_func}\\
		\notag
		&\qquad \textrm{s.t. }\\
		&{x}_{k+1,j} =\mathcal{F}({x}_{k,j},u_{k,j}^{\text{irr}},\hat{u}_k,\bm{{\phi}}_j), &&j\in [1,M], ~&&k\in [d,d+N-1]   \label{eq:re_a}\\
		&y_{k,j} = \mathcal{H}(x_{k,j},\bm{{\phi}}_j)&&j\in [1,M] ,~&&k \in [d+1,d+N]\label{eq:re_b}\\
		&\theta^{\text{RZ}}_{k,j} = \mathcal{M}(y_{k,j}) &&j\in [1,M] ,~&&k \in [d+1,d+N]\label{eq:re_c}\\
		&{x}_{d,j} = \hat{x}_j(d),  &&j\in [1,M]&& \label{eq:initial_state}\\
		& \underline{\nu}_j - \underline{\epsilon}_{k,j} \leq \theta^{\text{RZ}}_{k,j}\leq \bar{\nu}_j + \bar{\epsilon}_{k,j}, &&j\in [1,M] ,&&k \in [d+1,d+N]\label{eq:zone_b}\\
		&c_{k} \underline{u}_{j}^{\text{irr}}\leq u^{\text{irr}}_{k,j}\leq c_{k} \bar{u}_{j}^{\text{irr}},&& j\in [1,M] ,&&k \in [d,d+N-1]\label{eq:irrig_rate2}\\
		&c_{k}\in \{0, 1\}, && k \in [d,d+N-1]&&\\
		&\underline{\epsilon}_{k,j} \geq 0, \bar{\epsilon}_{k,j}\geq 0, &&j\in[1,M] ,&&k \in [d+1,d+N] \label{eq:lastconst_sv}
	\end{alignat}
\end{subequations}

where $\bm{\bar{E}} \coloneqq [\bm{\bar{\epsilon}}_1,\bm{\bar{\epsilon}}_2,...,\bm{\bar{\epsilon}}_M ]$, $\bm{\underline{E}} \coloneqq [\bm{\underline{\epsilon}}_1,\bm{\underline{\epsilon}}_2,...,\bm{\underline{\epsilon}}_M ]$, $\bm{U} \coloneqq [\bm{u}_1^{\text{irr}},\bm{u}_2^{\text{irr}},....,\bm{u}_M^{\text{irr}}]$, $\bm{\bar{\epsilon}}_{(\cdot)}\coloneqq [ \bar{\epsilon}_{d+1}, \bar{\epsilon}_{d+2},...,\bar{\epsilon}_{d+N}]$, $\bm{\underline{\epsilon}}_{(\cdot)}\coloneqq [ \underline{\epsilon}_{d+1}, \underline{\epsilon}_{d+2},...,\underline{\epsilon}_{d+N}]$, $\bm{c}\coloneqq [ c_{d}, c_{d+1},...,c_{d+N-1}]$, $\bm{u}^{\text{irr}}_{(\cdot)}\coloneqq [ u_{d}^{\text{irr}}, u_{d+1}^{\text{irr}},...,u_{d+N-1}^{\text{irr}}]$.
In $\mathbb{P}_{N,M}(d)$, $\underline{\epsilon}_{(\cdot)}$ and $\bar{\epsilon}_{(\cdot)}$ are slack variables that are introduced to relax the target zone $(\underline{\nu}_{(\cdot)},~\bar{\nu}_{(\cdot)})$.  For a MZ $j$, the maximum and minimum irrigation amounts that can be supplied during a particular irrigation event are denoted with $\bar{u}_{(\cdot)}^{\text{irr}}$ and $\underline{u}_{(\cdot)}^{\text{irr}}$, respectively. The binary variable $c_k$ is used in $\mathbb{P}_{N,M}(d)$ to encode the irrigation timing, indicating whether the irrigation event should be performed on day 
$k$ within $N$. Equation~(\ref{eq:irrig_rate2}) ensures that the irrigation timing is applied uniformly across the $M$ MZs that make up the field, in accordance with the daily irrigation scheduling problem.
The terms $\underline{Q}_{(\cdot)}$ and $\bar{Q}_{(\cdot)}$ are the per-unit costs associated with the violation of the lower and upper bounds of the target zone, respectively. The term $R_c$ is the fixed cost associated with the operation of the irrigation system, and $R_u$ is the per-unit cost of the irrigation amount $u^{\text{irr}}$.
The term $\hat{u}_k$ denotes the uncontrollable inputs (namely the reference evapotranspiration, rain, crop coefficient and rooting depth) for day $k$, while $\mathcal{M}(\cdot)$ is a linear function used to calculate the root zone soil moisture content ($\theta^{\text{RZ}}$) in each MZ. This calculation considers the spatial distribution of soil moisture ($y$) in the soil column of each MZ. The weighting approach applied assigns 40\% weight to the average volumetric moisture content in the upper quarter of the rooting depth ($z_{r}$), 30\% to the second quarter, 20\% to the third quarter, and 10\% to the last quarter. Note that this weighting approach reflects the relative importance of moisture at different depths within the rooting depth. According to Equation~(\ref{eq:initial_state}), the initial states, in each MZ, required for the evaluation of $\mathbb{P}_{N,M}(d)$, are estimated. Specific details concerning the estimation of the initial soil moisture content are provided in Section~\ref{sec:case_study_1}.

The terms $\bar{\nu}_j$ and $\underline{\nu}_j$ (also known as the threshold volumetric moisture content $\theta^{\text{th}}_j$) are calculated as follows:
\begin{align}\label{eq:zone_bounds}
	\bar{\nu}_j &= \theta^{\text{FC}}_j\\
	\underline{\nu}_j &=\theta^{\text{TH}}_j =  \theta^{\text{FC}}_j - \left[ \text{MAD}\times\left(\theta^{\text{FC}}_j -  \theta^{\text{WP}}_j\right)\right]
\end{align}
where $\theta^{\text{FC}}_j$ is the volumetric moisture content at field capacity for MZ $j$, $\theta^{\text{WP}}_j$ represents the volumetric water content at the wilting point for MZ $j$, and MAD refers to the management allowable depletion, which indicates the fraction of the total available water that is permitted to be depleted. 
\section{Enhancing the Computational Efficiency of $\mathbb{P}_{N,M}(d)$ Using a ReLU Neural Network} \label{relu_mod_dev}

While  $\mathbb{P}_{N,M}(d)$ provides an optimal control framework for addressing the daily irrigation scheduling problem comprehensively, it presents significant computational challenges. These challenges primarily stem from the inherent complexity of mixed-integer problems. Moreover, using the Richards equation to represent soil moisture dynamics in 
$\mathbb{P}_{N,M}(d)$ adds further complexity. The Richards equation is a highly nonlinear partial differential equation, and its nonlinearity, combined with its dependence on complex soil hydraulic functions, makes it computationally expensive to solve, particularly when integrated into optimization frameworks. Additionally, the numerical solution of the Richards equation requires discretization in both time and space, transforming it into a large system of algebraic equations. This discretization increases the problem's dimensionality. These complexities are further compounded when $\mathbb{P}_{N,M}(d)$ is applied to large-scale fields with multiple MZs, where numerous variables must be considered simultaneously.

Past studies seeking to enhance the computational efficiency of 
$\mathbb{P}_{N,M}(d)$ have explored replacing the Richards equation with machine learning surrogate models like LSTM networks.

To further enhance the computational efficiency of 
$\mathbb{P}_{N,M}(d)$, this work proposes using a feedforward ReLU neural network to model soil moisture dynamics in each MZ of the field, instead of machine learning surrogate models like the LSTM network, for two main reasons. First, compared to the Richards equation, which is highly nonlinear and requires complex functions for its numerical solution, the ReLU network is a simpler model. Therefore, replacing the Richards equation with a ReLU neural network in 
$\mathbb{P}_{N,M}(d)$ can potentially reduce the solution time.

Second, a feedforward ReLU neural network can be exactly represented using MIL constraints. Incorporating the MIL constraint representation of the ReLU neural network within 
$\mathbb{P}_{N,M}(d)$ further enhances computational efficiency. Notably, the cost function of 
$\mathbb{P}_{N,M}(d)$ is quadratic, and, except for the constraints describing soil moisture dynamics across the MZs (i.e.  Equations~(\ref{eq:re_a}) and ~(\ref{eq:re_b})), all other constraints in 
$\mathbb{P}_{N,M}(d)$ are linear. Therefore, adopting a MIL representation of the ReLU neural network transforms the problem into a MIQP problem, which is the simplest class of MINLP problems. The resulting MIQP problem can then be solved using solvers such as Gurobi.

Based on the foregoing, this section describes how feedforward ReLU neural networks, which model the soil moisture dynamics in the various MZs of the field, can be trained and integrated into 
$\mathbb{P}_{N,M}(d)$ to enhance its computational efficiency. The section begins with an overview of ReLU neural networks and the MIL formulation of these networks. It then discusses the training process for a ReLU neural network tailored to each MZ of the field. Details regarding the proposed neural network's design, the generation of training data, and the evaluation of the trained networks are also provided.

\subsection{MIL representation of ReLU Neural Networks}\label{sec:mil_section}
A ReLU neural network is a feedforward neural network where the ReLU activation function is applied to the neurons, except for the neurons in the output layer. Mathematically, a ReLU neural network can be represented as follows:
\begin{subequations}\label{eq:relu_nn}
	\begin{align}
		\mathcal{Z}_0 &= \mathcal{X}\\
		\mathcal{Z}_{\ell + 1} &= \text{max}\left( W_{\ell} \mathcal{Z}_{\ell} + b_{\ell}, 0 \right),\quad \ell = 0, 1,...,L-1\\
		\mathcal{F}_{\text{NN}}(\mathcal{X}) &= W_L\mathcal{Z}_L + b_L
	\end{align}
\end{subequations}

where $\mathcal{Z}_0 \in \mathbb{R}^{n_{0}}$ is the input to the ReLU neural network, $\mathcal{Z}_{\ell+1}\in \mathbb{R}^{n_{\ell +1}}$ is the output vector of the $(\ell +1)$-th hidden layer, which has $n_{\ell + 1}$ neurons. The output of the network is denoted with $\mathcal{F}_{\text{NN}}(\mathcal{X}) \in \mathbb{R}^{n_{L +1}} $, and $W_{\ell} \in \mathbb{R}^{n_{\ell +1}\times n_{\ell}}$, $b_{\ell} \in \mathbb{R}^{n_{\ell +1}}$ are the weight and bias vectors of the ($\ell+1$)-th hidden layer.

For the ($\ell+1$)-th hidden layer in the ReLU neural network described in Equation~(\ref{eq:relu_nn}), let $\underline{U}^{\ell}$ and $\bar{U}^{\ell}$ be the element-wise lower and upper bounds on the input i.e. $\underline{U}^{\ell}\leq W_{\ell} \mathcal{Z}_{\ell} + b_{\ell} \leq\bar{U}^{\ell}$. The ReLU activation function is equivalent to the following set of MIL constraints~\cite{tjeng2017evaluating}:

\begin{equation}\label{eq:mil_const}
	\mathcal{Z}_{\ell + 1} = \text{max}\left( W_{\ell} \mathcal{Z}_{\ell} + b_{\ell}, 0 \right)  \equiv  \begin{cases}
		\mathcal{Z}_{\ell + 1} \geq  W_{\ell} \mathcal{Z}_{\ell} + b_{\ell}\\
		\mathcal{Z}_{\ell + 1} \leq  W_{\ell} \mathcal{Z}_{\ell} + b_{\ell} - \text{diag}\left(\underline{U}^{\ell}\right)(1-e_{\ell})\\
		\mathcal{Z}_{\ell + 1} \geq 0 \\
		\mathcal{Z}_{\ell + 1} \leq \text{diag}\left(\underline{U}^{\ell}\right)(e_{\ell})
	\end{cases}
\end{equation}
where $e_{\ell} \in \{0,1\}^{n_{\ell+1}}$ is a vector of binary variables for the ($\ell+1$)-th activation layer. Given that input to the ReLU neural network is bounded, the interval bound propagation approach~\cite{cheng2017maximum} is used to determine the tight bounds $\left[\underline{U}^{\ell}, \bar{U}^{\ell}\right]$ on the input to each ReLU unit, so as to reduce numerical issues and enhance solution times.

\subsection{ReLU Network Development}
This section details the development of a ReLU neural network for modeling soil moisture dynamics across each MZ of the field. It begins with an overview of the ReLU neural network architecture designed for each MZ, followed by a discussion on the generation of the training data used for neural network training. The section concludes with a description of the methodology employed to evaluate the predictive performance of the trained neural network.

\subsubsection{ReLU Neural Network Design}\label{sec:relu_design}
For each MZ, a ReLU network is trained to make daily predictions of the root zone volumetric moisture content. In this work, the universal approximation property of neural networks is leveraged to directly model the root zone soil moisture (a scalar value), which is calculated as a weighted sum (i.e. the linear function $\mathcal{M}(\cdot)$ in $\mathbb{P}_{N,M}(d)$) of the soil moisture contents at various depths in the soil column. This approach differs from existing soil moisture surrogate modeling methods, such as the one presented in~\cite{gu2021neural}, where multiple neural networks are trained to predict soil moisture dynamics at different depths, and the root zone soil moisture is subsequently computed as a weighted sum of these individual predictions.  The proposed ReLU neural network approach offers a more compact model that can further enhance the computational efficiency of $\mathbb{P}_{N,M}(d)$.

In particular, the ReLU neural network is trained to make one-day-ahead root zone soil moisture content predictions. To make these predictions, it employs both current ($k$) and past inputs ($k-1$, $k-2$, ..., $k-l$), where $l$ represents the time lag (in days). The specific inputs to the network include the  root zone volumetric water content ${\theta}^{\text{RZ}}$, precipitation $I$ (the sum of rain $R$ and irrigation amount), crop coefficient $\text{K}_{\text{c}}$, reference evapotranspiration $\text{ET}_0$, and the rooting depth $\text{z}_{\text{r}}$. 
\begin{equation}\label{eq:ReLU}
{\theta^{\text{RZ}}}_{k+1} = \mathcal{N}(\{{\gamma}\}_{k-l}^{k},\eta)
\end{equation}
where $\{\gamma\}_{k-l}^{k} \coloneqq [ \gamma_{k-l}, \gamma_{k-l-1},..,\gamma_{k}]$, $\gamma\in[{\theta}^{\text{RZ}},\text{K}_{\text{c}}, \text{ET}_0, I, \text{z}_{\text{r}}]$ and $I=u^{\text{irr}} + R$. In Equation~(\ref{eq:ReLU}), $\eta$ is a compact representation of the weights and biases of the ReLU neural network $\mathcal{N}$. 

Based on Equation~(\ref{eq:ReLU}), the input vector to the ReLU network contains $5(l+1)$ elements, and its output contains 1 element. Thus, the terms $n_0$ and $n_{L+1}$ defined in Section~\ref{sec:mil_section} are equal to $5(l+1)$ and $1$, respectively.

Since the ReLU neural network directly models the root zone soil moisture content in each zone, Equation~(\ref{eq:ReLU}) approximates and replaces Equations~(\ref{eq:re_a}),~(\ref{eq:re_b}), and~(\ref{eq:re_c}) in $\mathbb{P}_{N,M}(d)$. Furthermore, Equation~(\ref{eq:mil_const}) is then used to represent  Equations~(\ref{eq:ReLU}) within $\mathbb{P}_{N,M}(d)$.

\subsubsection{Training Data Generation}
In this study, open-loop simulated data from the calibrated 1D Richards equation for each MZ are used to train a ReLU neural network specific to that MZ. The calibration process involves using soil moisture observations from the field to estimate the soil hydraulic parameters, which are then employed to calibrate the Richards equation. This calibration ensures that the resulting Richards equation accurately describes the soil moisture dynamics of the field. 

For a carefully chosen initial state---specifically, the converged soil moisture estimates from an offline data assimilation process in this work---the training dataset is generated by solving Equations~(\ref{eq:state_equation}) and (\ref{eq:output_equation}) for randomly generated input $u_k$ trajectories.  Noise is  included in the open-loop simulations to account for model uncertainty and to enhance the robustness of the neural network. The use of randomly generated inputs is employed to ensure that the resulting time series dataset encompasses the dynamics of soil moisture under a wide-range of conditions.

Note that since the ReLU neural network is trained to directly predict the root zone soil moisture content, the function  $\mathcal{M}$ outlined in formulation $\mathbb{P}_{N,M}(d)$, together with the spatial volumetric moisture content at each time instant, is used to calculate the corresponding root zone soil moisture content for the training dataset before the ReLU neural network is trained.

\subsubsection{ReLU Neural Network Evaluation}
To evaluate the predictive performance of the trained ReLU neural network, a comparison with the calibrated Richards equation is conducted. In this study, a specific initial state is chosen, and predetermined trajectories of the inputs  $u_k$ are defined for a given period.

Both the identified ReLU neural network and the calibrated 1D Richards equation are then simulated using the selected initial state and predetermined input trajectories. The predictions from the ReLU neural network are compared to the root zone soil moisture content computed by the calibrated Richards equation, using the root mean squared error (RMSE) metric.

\section{Study Area}\label{sec:study_area}

\begin{figure}[!ht]
	\subfloat[Study Area]{
		\begin{minipage}[c]{0.50\columnwidth}
			\centering
			\includegraphics[width=\columnwidth]{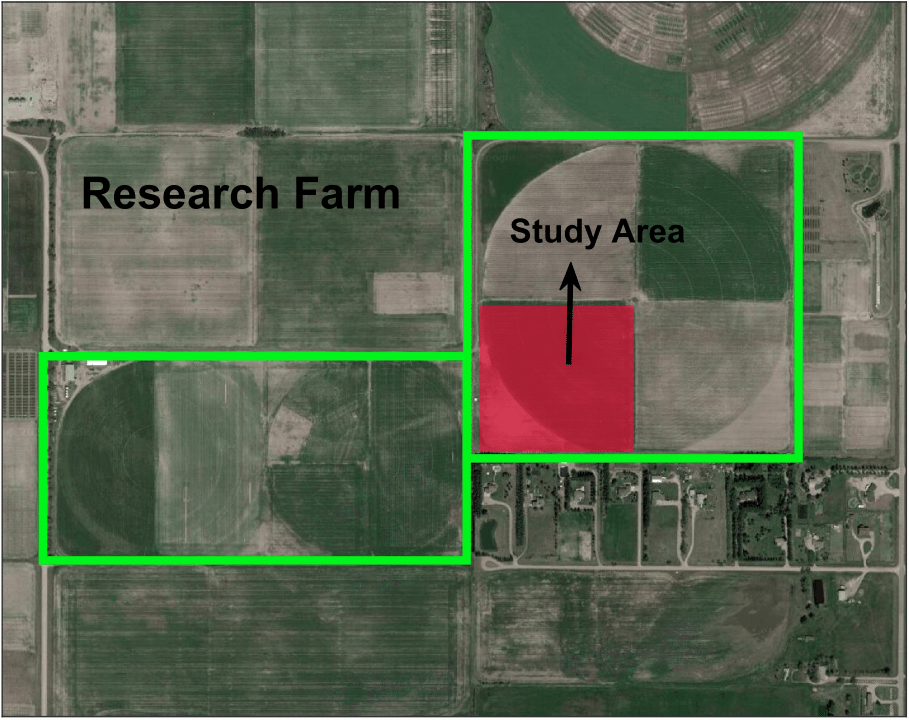}
	\end{minipage}}
	\hfill
	\hspace{4mm}
	\subfloat[MZ Map.]{
		\begin{minipage}[c]{0.47\columnwidth}
			\centering
			\includegraphics[width=\columnwidth]{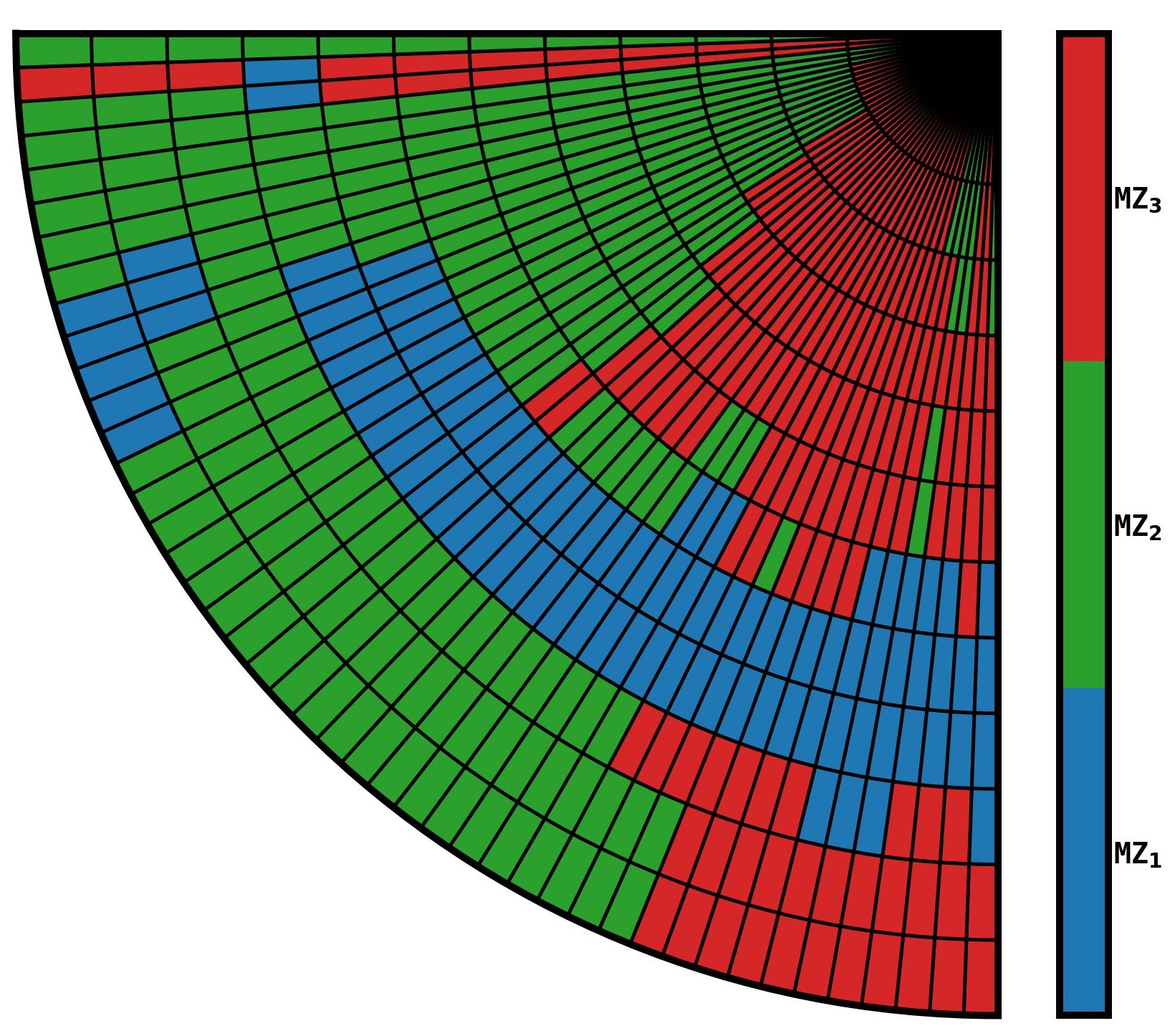}
	\end{minipage}}
	\caption{Study area and its management zone map.}
	\label{fig:study_area_mzd}
\end{figure}
The mixed-integer MPC scheduler with ReLU neural network was applied to a specific quadrant of a large-scale circular field, highlighted by the red rectangle in Figure~\ref{fig:study_area_mzd}(a). This study area is located at a Research Farm in Lethbridge, Alberta, Canada. The field is equipped with a variable-rate center pivot irrigation system, spanning a length of 294 meters, to facilitate irrigation management. The application considered that spring soft wheat was cultivated in the study area.

Before applying the scheduler, the investigated quadrant was delineated into MZs using a three-stage delineation method originally proposed in \cite{agyeman2024learning}. This method incorporated elevation and soil hydraulic parameters, utilizing the k-means clustering technique for MZ delineation. The soil hydraulic parameter attributes were derived from an offline data assimilation method, which integrated remotely sensed soil moisture observations into the cylindrical coordinate version of the Richards equation using the extended Kalman filter. For specific details on the data assimilation approach, interested readers may refer to~\cite{agyeman2022simultaneous}. Figure~\ref{fig:study_area_mzd}(b) shows the three distinct MZs identified in the investigated quadrant.

\subsection{Surrogate Modeling of the Study Area}
Prior to generating the open-loop data for training the 3 ReLU neural networks for $\mathbb{P}_{N,M}(d)$, the soil moisture dynamics in each of the three MZs of the study area were modeled using the 1D Richards equation. For each MZ, the 1D Richards equation was calibrated with the centroidal soil hydraulic parameters specific to that zone. Subsequently, extensive open-loop simulations of these calibrated Richards equations were conducted, incorporating randomly generated inputs. 

The neural network training stage incorporated two different rooting depths: 0.5 m and 1.00 m, which is consistent with irrigation management for spring soft wheat. The reference evapotranspiration was randomly generated within the ranges of 0.1 mm/day to 8.99 mm/day for all three MZs. The irrigation amount was randomly generated between 4.0 mm/day and 52.0 mm/day for MZ$_1$, 4.3 mm/day and 59.6 mm/day for MZ$_2$, and 5.0 mm/day and 62.3 mm/day for MZ$_3$. Notably, the selection of the irrigation amount  and reference evapotranspiration ranges was informed by irrigation management practices on the study area and historical weather, respectively.

Rain was randomly selected from historical rainfall data from the 2005 to 2014 growing seasons.
The crop coefficient data were generated using the empirical formula in Equation~\ref{eq:kc_relation}, along with randomly generated average temperature values from the same period. The simulation settings used for these open-loop simulations are detailed in Table~\ref{tbl:sim_settings} of Appendix~(\ref{sec:settings_RE}). 

The ReLU neural networks were trained using the Keras library in Python. The model architecture and hyperparameters employed during the training process are provided in Table~\ref{tbl:hyper_pars} of Appendix~(\ref{sec:lstm_network}). These values were determined through simulation experiments; thus, it is probable that improved results could be achieved through a systematic tuning of the hyperparameters.

\section{Performance Evaluation}
Two main studies were conducted to assess the performance of the mixed-integer MPC with the ReLU neural network scheduler. In the first study, referred to as Case Study 1, the computational benefits of employing the ReLU neural network were investigated. In this study, the mixed-integer MPC with the ReLU neural network was used to provide irrigation schedules for the 2015 and 2022 growing seasons, which were specifically chosen to assess the schedulers under relatively dry and wet conditions, respectively. The reference evapotranspiration and the average temperature trajectories for the 2015 and 2022 growing seasons are depicted in Figure~\ref{fig:weather_data} of Appendix~\ref{section:weather_data}.

Similarly, the mixed-integer MPC incorporating the LSTM network, originally proposed by~\cite{agyeman2023lstm}, was also employed to generate schedules for the same growing seasons. This approach was chosen for two key reasons: first, the LSTM-based framework was found to significantly enhance computational efficiency compared to the use of the Richards equation $\mathbb{P}_{N,M}(d)$; second, the mixed-integer MPC with LSTM demonstrated better performance than the widely used triggered irrigation scheduling approach in terms of water savings and enhancing optimal crop yield.
 The two approaches were compared in terms of the average solution time of 
$\mathbb{P}_{N,M}(d)$. Additionally, the approaches were compared based on total prescribed irrigation and Irrigation Water Use Efficiency (IWUE), defined as the ratio of predicted crop yield to total prescribed irrigation.

In the second study, named Case Study 2, the water-saving benefits and the ability of the mixed-integer MPC with the ReLU neural network to enhance IWUE were evaluated. This case study was performed to compare the ReLU-based approach against the widely used triggered irrigation scheduling method, providing insights into the practical applicability of the ReLU-based scheduler in a real-world context. Case Study 2 was also conducted over the 2015 and 2022 growing seasons. The two approaches were compared in terms of total prescribed irrigation and IWUE to determine which method is more effective in conserving water while maintaining or improving crop yield. It is important to note that in both Case Studies 1 and 2, a MAD value of 50\% was used, consistent with the irrigation management practices for spring soft wheat. 

The subsequent sections will discuss the settings under which these two case studies were conducted.

\subsection{Simulation Settings of Case Study 1}\label{sec:case_study_1}
Table~\ref{tbl:parameter_values} of Appendix~(\ref{sec:tuning_parameters}) provides the relevant parameters for the formulation $\mathbb{P}_{N,M}(d)$ used in this case study. It is important to note that the per-unit costs were treated as tuning parameters in this study. 

In the mixed-integer MPC with LSTM models, the LSTM models were trained according to the approach proposed in~\cite{agyeman2024learning}. Specific details for training the LSTM networks are omitted due to space constraints; interested readers may refer to~\cite{agyeman2024learning} for the specific training details. It is important to note that the same training data were used to train both the LSTM and ReLU neural network models for the three MZs of the study area.  Additionally, the ReLU neural network design outlined in the Section~\ref{sec:relu_design} was adopted for the training of the LSTM networks for each MZ. However, the tuning of hyperparameters and the determination of the time lag $l$ for the training of the LSTM and ReLU networks were conducted independently, as each network has different characteristics for achieving optimal performance.

In both scheduling approaches considered in this study, the simulation period spanned from May 5th to September 4th, for the 2015 and 2022 seasons. Throughout the evaluation process, a rooting depth of 0.5 m was considered from May 5th to July 15th, after which a rooting depth of 1.0 m was used until the end of the growing season. This trajectory of the rooting depth aligns with standard irrigation management for the soft spring wheat crop. A prediction horizon (and control horizon) of 7 days was employed during the evaluation of the two scheduling approaches. Several simulation experiments demonstrated that a 7-day prediction horizon effectively captures the essential dynamics of the study area. This period also allows the scheduler to utilize the most accurate weather forecasts available.

The initial root zone soil moisture content in the three MZs of the field, required for evaluating the two scheduling approaches, was not assumed to be known. Instead, the study simulated the presence of a remote sensor in each MZ, reflecting the actual field conditions where microwave remote sensors are mounted. These sensors provided daily soil moisture observations corresponding to the average moisture content in the top 25 cm of the soil column. The soil moisture content was estimated from these daily observations using the EKF, which was chosen due to its demonstrated effectiveness in accurately estimating soil moisture content in agro-hydrological systems~\cite{medina2014kalman,lu2011dual}. The specific EKF design for each MZ is detailed in  Appendix~\ref{sec:ekf_design}.

On the first day of the evaluation period, the two scheduling schemes were evaluated using the initial guess of the soil moisture distributions for the various MZs, along with the 7-day weather and crop information predictions. Although the actual weather information for the growing seasons was known during this simulation, uncertainty was incorporated into the weather forecasts used in evaluating the agents and in the MPC. This uncertainty was modeled as a normal distribution with a mean of 0 and a specified standard deviation. As the prediction horizon extended further into the future, the standard deviation values were gradually increased to reflect the growing uncertainty associated with longer-term weather predictions. The mixed-integer MPC with ReLU neural networks was solved using the Gurobi solver, while the mixed-integer MPC with LSTM networks was solved using the BONMIN solver. Additionally, the OMLT package~\cite{ceccon2022omlt} in Python was used to translate the trained ReLU neural networks into the MIL constraints, and the resulting MIQP was solved in Pyomo~\cite{hart2017pyomo}.

The irrigation amounts obtained from solving the two MPCs were implemented in a receding horizon fashion, where the first irrigation amount was applied to the actual field, and the rest were discarded. The actual field conditions were represented using three well-calibrated 1D Richards equations. The initial soil moisture contents in the MZs of the actual field were initialized with the converged soil moisture estimates obtained from the offline state and parameter estimation outlined in Section~\ref{sec:study_area}. On the second day, measurements corresponding to the average soil water content in the top 25 cm were obtained for each MZ from the actual field. To account for sensor noise, noise from a Gaussian distribution with a mean of 0 and a standard deviation of 0.0008 was added to these measurements, reflecting the measurement noise observed in the microwave radiometers mounted on the study area. These noisy measurements were combined with the current soil moisture estimates in each MZ using the EKF to update the soil moisture estimates. The updated soil moisture estimates, along with the forecasted weather and crop coefficient data, were then used to prescribe the irrigation timing and amounts for the MZs. This process was repeated daily until the end of the investigation period for the two growing seasons.

The predicted yield of soft spring wheat, required for the calculation of the IWUE under the two scheduling schemes, was calculated according to the equations outlined in Appendix~\ref{sec:predicted_yield}.

\subsection{Simulation Settings for Case Study 2}
Before describing the settings under which this study was performed, a brief description of the triggered irrigation scheduling approach is presented.

The triggered approach aims to maintain the root zone soil moisture within the lower and upper bounds of the target zone ($\underline{\nu}$, $\bar{\nu}$). For MZ $j$, an irrigation event is triggered on a day $d$, when the daily root zone moisture content $\theta^{\text{RZ}}_d$ falls below the lower bound of the target zone. When this occurs, the irrigation amount $u^{\text{irrig}}_{d}$ is calculated using the following formula~\cite{gu2021neural}:
\begin{equation}
	u^{\text{irrig}}_{d} = \left[(\bar{\nu} - \theta^{\text{RZ}}_d) \times \text{z}_{\text{r}}\right] - P_{d+4}
	\label{eq:irrig_rate}
\end{equation}
where $P_{d+4}$ indicates the cumulative rainfall that occurs from day $d+1$ to $d+4$.

The implementation of the mixed-integer MPC with the ReLU neural network in this case study is the same as its evaluation in Case Study 1. Therefore, this section focuses on the evaluation settings of the triggered scheduling approach for the 2015 and 2022 seasons.

For both growing seasons, the implementation of the triggered approach spanned from May 5th to September 4th. The rooting depth trajectory, outlined in Section~\ref{sec:case_study_1} is adopted in this study. On the first day of the evaluation period, the triggered  scheduling scheme was evaluated with the initial soil moisture distribution estimates for the various MZs, along with the 4-day rainfall prediction.
In each MZ, the current root zone soil moisture content was compared to the lower bounds of their target soil moisture range. If the root zone soil moisture content in all three MZs remained above their lower bounds, an irrigation amount of 0 was applied to the actual field. However, if the root zone soil moisture content fell below the lower bound in at least one MZ, a non-zero irrigation rate was applied to all MZs. These non-zero irrigation rates were calculated using Equation~(\ref{eq:irrig_rate}), for all three MZs. 

During the implementation of the triggered approach, the actual field conditions used in Case Study 1 were employed. Additionally, the soil moisture content in each MZ was estimated using the same approach outlined in Case Study 1, with the estimator design for each MZ detailed in Appendix~\ref{sec:ekf_design}.

\section{Results and Discussion}
In this section, the results of the case studies are presented and analyzed in detail. The section begins with a discussion of the outcomes from the validation experiment conducted on the trained ReLU neural networks. This is followed by a presentation and analysis of the results comparing the mixed-integer MPC with ReLU neural networks to the mixed-integer MPC with LSTM networks. Finally, the section concludes with a detailed discussion of the results comparing the mixed-integer MPC with ReLU neural networks to the triggered irrigation scheduling approach. 

\subsection{Predictive Performance of the Identified ReLU Neural Networks}
\begin{figure}[!ht]
	\centerline{\includegraphics[width=0.75\textwidth]{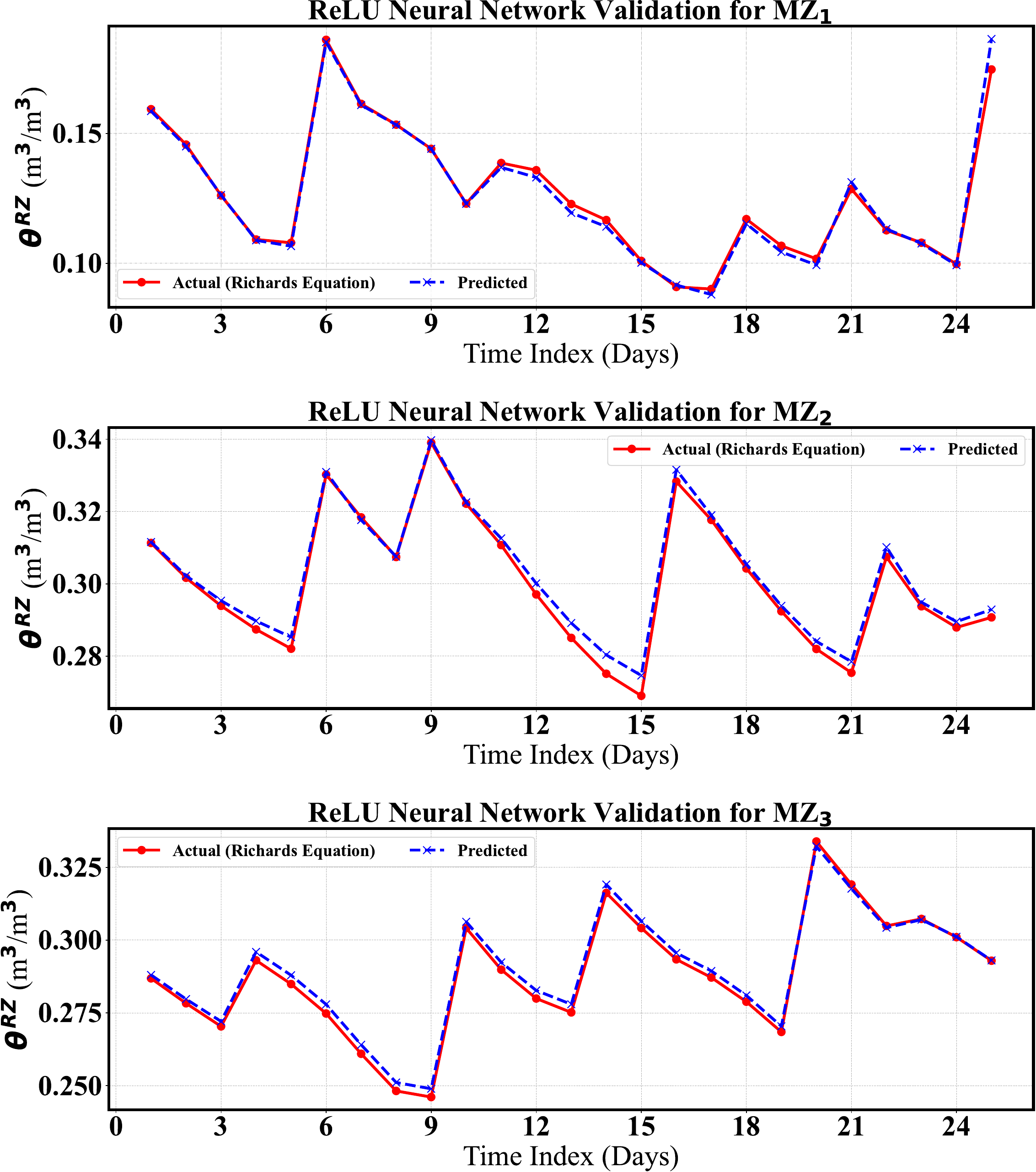}}
	\caption{Predictive performance of the identified ReLU neural networks.} 
	\label{fig:predictive_perform}
\end{figure}

The results of the evaluation of the predictive performance of the ReLU networks, conducted over a 25-day period, are depicted in Figure~\ref{fig:predictive_perform}.
From this figure, the soil moisture predictions can be considered accurate, as there is a strong agreement between the predicted root zone soil moisture content and the actual values derived from the calibrated 1D Richards equation. This accuracy is further confirmed by the low RMSE values shown in Table~\ref{tbl:predictive_comp}, which highlight the good predictive capacity of the ReLU neural networks for root zone soil moisture content across all three MZs.

The results demonstrate that the proposed ReLU neural network modeling approach for root zone soil moisture in the MZs of the study area is effective. These findings are consistent with other studies that have utilized ReLU neural networks to describe soil moisture dynamics in agro-hydrological systems~\cite{adeyemi2018dynamic,achieng2019modelling}.
Furthermore, these results suggest that the ReLU neural networks' capacity for universal approximation can be leveraged to directly model root zone soil moisture content, potentially enhancing computational efficiency when employed in schedulers that require the solution of a numerical optimization problem. This approach contrasts with earlier methods~\cite{gu2021neural}, where separate neural network models were trained to predict soil moisture content at different depths within the soil profile, and a weighted approach was subsequently used to estimate root zone soil moisture content from these predictions.

The results obtained from the validation experiment further highlight the accuracy of the identified ReLU networks in performing multi-step-ahead predictions. In this experiment, the ReLU networks, originally trained to  perform one-step-ahead predictions, were employed recursively over a 25-day period. This recursive strategy involved using the output from one time step as an input for the next. This property is particularly desirable in MPC frameworks because MPC relies on the ability to make accurate predictions over the prediction horizon to determine optimal control actions.

\begin{table}[t]
	\caption{Predictive performance of the trained ReLU Neural Networks over a period of 25 days.}
	\small
	\centering
	\begin{tabular}{cc}
		\toprule
		& \textbf{RMSE} \\
		\midrule
		MZ$_1$ & 0.0028 \\
		MZ$_2$ & 0.0025 \\
		MZ$_3$ & 0.0022 \\
		\bottomrule
	\end{tabular} \label{tbl:predictive_comp}
\end{table}

\subsection{Comparison of $\bm{\mathbb{P}_{N,M}(d)}$ with ReLU and $\bm{\mathbb{P}_{N,M}(d)}$ LSTM Approaches}

Tables~\ref{tbl:lstm_relu_dry} and~\ref{tbl:lstm_relu_wet} provide a summary of the comparison between the ReLU and LSTM approaches in terms of total irrigation, IWUE, and average solution time for the 2015 and 2022 growing seasons.

For the 2015 season (Table~\ref{tbl:lstm_relu_dry}), which represents a relatively dry period, the total irrigation required for both the ReLU and LSTM approaches is almost identical (0.983 m for ReLU vs. 0.984 m for LSTM). The IWUE is slightly higher for the LSTM approach (8.84 kg/m$^3$) compared to the ReLU approach (8.63 kg/m$^3$). However, the ReLU neural network shows a significant computational advantage, reducing the average solution time of $\mathbb{P}_{N,M}(d)$ from 107.5 minutes (LSTM) to 0.44 minutes (ReLU), which represents a reduction of approximately 99.6\%.

For the 2022 season (Table~\ref{tbl:lstm_relu_wet}), which represents a relatively wet period, the total irrigation required is slightly lower for the ReLU approach (0.761 m) compared to the LSTM approach (0.790 m). The IWUE for both approaches is very similar, with 10.65 kg/m$^3$
for ReLU and 10.71 kg/m$^3$ for LSTM. The ReLU approach again shows a significant reduction in computational time, with an average solution time of 0.52 minutes compared to 82 minutes for the LSTM, translating to a reduction of approximately 99.4\%.

These results confirm that exploiting ReLU neural networks and their MIL representation significantly enhances the computational efficiency of the mixed-integer MPC-based irrigation scheduling. The ReLU-based approach achieves a significant reduction in solution time---around 99.6\% and 99.4\% for the 2015 and 2022 seasons, respectively---while maintaining comparable levels of  total prescribed irrigation and IWUE. This significant improvement in computational efficiency makes the ReLU-based model a more suitable  option for real-time irrigation scheduling without compromising irrigation performance.

It has been reported in~\cite{grimstad2019relu} that there exist MIL representations for sigmoid and hyperbolic activation functions, albeit in an inexact manner. Since the LSTM network primarily comprises sigmoid and hyperbolic activation functions, an inexact MIL approximation can be employed to represent the LSTM network in 
$\mathbb{P}_{N,M}(d)$. However, due to the inexact nature of this representation, the accuracy and computational efficiency of the resulting formulation may also be compromised compared to the exact MIL formulation for ReLU networks.

Despite this, the ReLU approach remains an attractive alternative over the LSTM approach. This is due to the fact that training a ReLU neural network has been reported to be more efficient from a computational standpoint compared to training an LSTM network~\cite{su2017probabilistic}. However, future studies will be needed to compare the performance of the MIL representations of LSTM and ReLU neural networks in the irrigation scheduling setting to better understand their relative strengths and limitations.

\begin{table}[t]
	\caption{Comparison between the $\mathbb{P}_{N,M}(d)$ with LSTM and $\mathbb{P}_{N,M}(d)$ with ReLU neural network for the 2015 season.}
	\small
	\centering
	\begin{tabular}{ccc}
		\toprule
		\textbf{} & \textbf{$\bm{\mathbb{P}_{N,M}(d)}$ with LSTM}& \textbf{$\bm{\mathbb{P}_{N,M}(d)}$ with ReLU}\\
		\midrule
		Total irrigation (m) & 0.984  & 0.983\\  
		IWUE (kg/m$^3$)& 8.84 & 8.63\\ 
		Average solution time of $\mathbb{P}_{N,M}(d)$ (minutes)& 107.5 & 0.44\\  \bottomrule 
	\end{tabular} \label{tbl:lstm_relu_dry}
\end{table}

\begin{table}[t]
	\caption{Comparison between the $\mathbb{P}_{N,M}(d)$ with LSTM and $\mathbb{P}_{N,M}(d)$ with ReLU neural network for the 2022 season.}
	\small
	\centering
	\begin{tabular}{ccc}
	\toprule
	\textbf{} & \textbf{$\bm{\mathbb{P}_{N,M}(d)}$ with LSTM}& \textbf{$\bm{\mathbb{P}_{N,M}(d)}$ with ReLU}\\
	\midrule
	Total irrigation (m) & 0.790  & 0.761\\  
	IWUE (kg/m$^3$)& 10.71 & 10.65\\ 
	Average solution time of $\mathbb{P}_{N,M}(d)$ (minutes)& 82 & 0.52\\  \bottomrule 
\end{tabular} \label{tbl:lstm_relu_wet}
\end{table}

\subsection{Comparison of $\bm{\mathbb{P}_{N,M}(d)}$ with ReLU and Triggered Approaches}

Tables~\ref{tbl:relu_trig_dry} and~\ref{tbl:relu_trig_wet} provide a comparison between the ReLU-based scheduling approach and the traditional triggered irrigation scheduling approach in terms of total irrigation and Irrigation Water Use Efficiency (IWUE) for the 2015 and 2022 growing seasons.

For the 2015 season (Table~\ref{tbl:relu_trig_dry}), which represents a relatively dry period with a total rainfall of 14.7 mm, the total irrigation required by the ReLU-based approach is 0.983 m, representing a reduction of approximately 9.8\% compared to the 1.09 m required by the triggered approach. The ReLU approach also demonstrates an enhanced IWUE of 8.63 kg/m$^3$, which is an improvement of approximately 14.6\% over the triggered approach's IWUE of 7.53 kg/m$^3$

For the 2022 season (Table~\ref{tbl:relu_trig_wet}), which represents a relatively wet period with a total rainfall of 230.9 mm, the ReLU-based approach requires 0.761 m of total irrigation, a reduction of about 21.6\% compared to the 0.97 m required by the triggered approach. The IWUE for the ReLU approach is 10.65 kg/m$^3$, reflecting an enhancement of approximately 36.9\% over the IWUE of 7.78 kg/m$^3$ for the triggered approach.

\begin{figure}[!ht]
	\centerline{\includegraphics[width=0.85\textwidth]{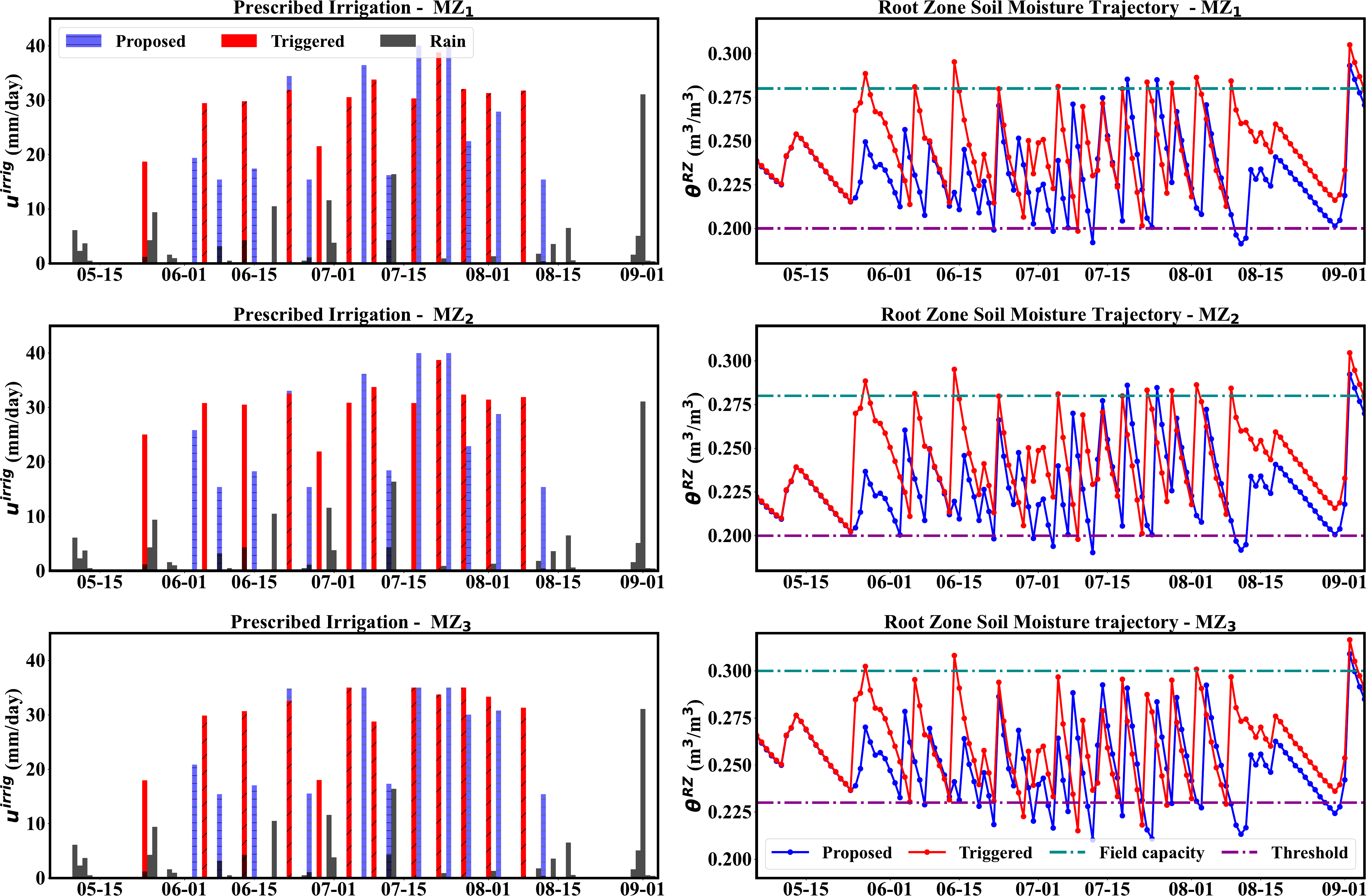}}
	\caption{Prescribed irrigation schedules and the trajectories of root zone soil moisture content for the 2015 season.} 
	\label{fig:dry}
\end{figure}

\begin{figure}[!ht]
	\centerline{\includegraphics[width=0.85\textwidth]{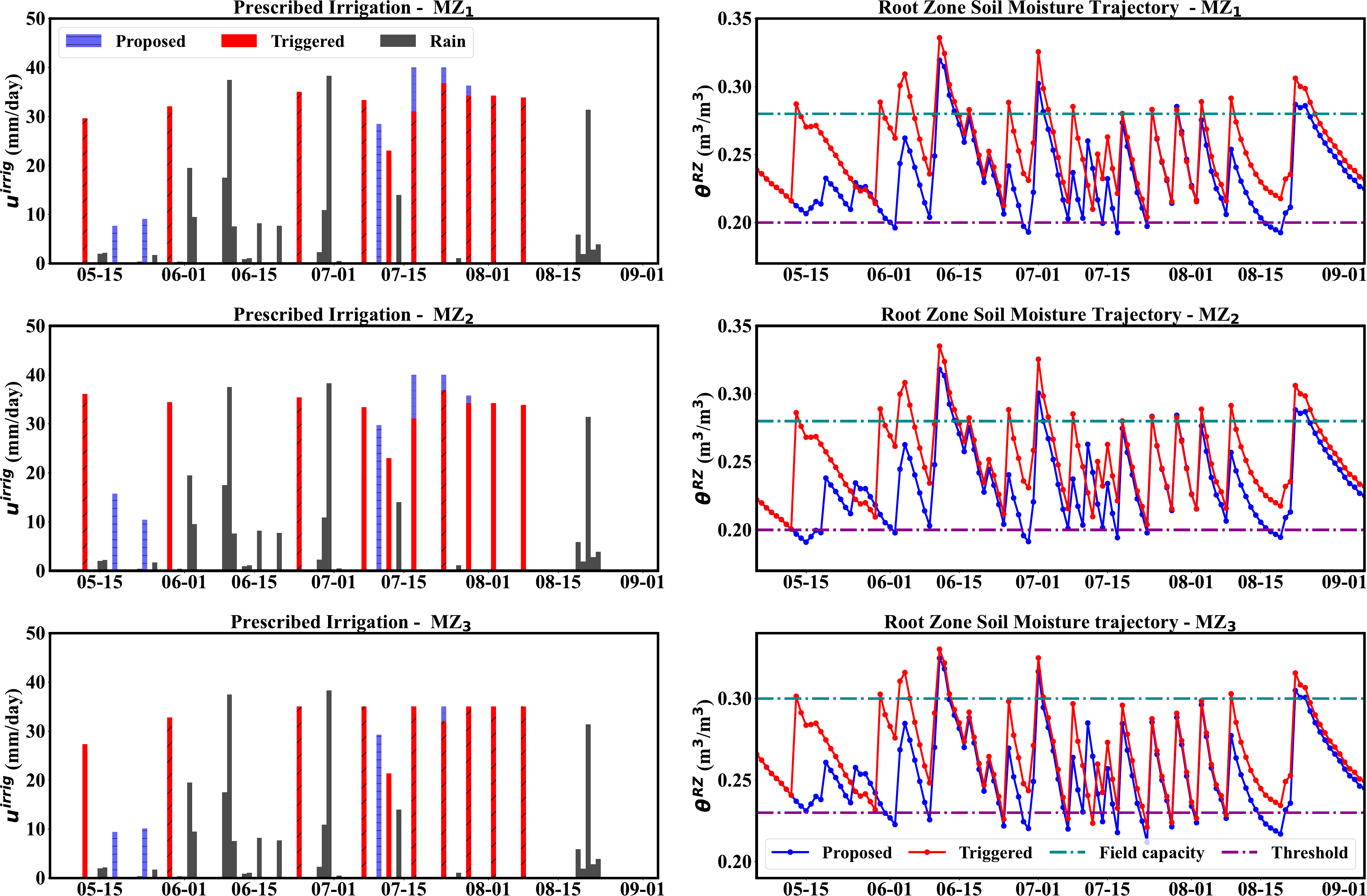}}
	\caption{Prescribed irrigation schedules and the trajectories of root zone soil moisture content for the 2022.} 
	\label{fig:wet}
\end{figure}

Additionally, as illustrated in Figures~\ref{fig:dry} and~\ref{fig:wet}, the ReLU-based approach more effectively maintains soil moisture levels within the optimal range across both seasons, with fewer and more controlled irrigation events compared to the triggered approach. 

These results, combined with the ReLU approach's significantly enhanced solution time, highlight its potential as an attractive option for real-time irrigation scheduling.

        \begin{table}[t]
	\caption{Comparison between the triggered and $\mathbb{P}_{N,M}(d)$ with ReLU neural network for the 2015 season.}
	\small
	\centering
	\begin{tabular}{ccc}
		\toprule
		\textbf{} & \textbf{$\bm{\mathbb{P}_{N,M}(d)}$ with ReLU}& \textbf{Triggered}\\
		\midrule
		Total irrigation (m) & 0.983  & 1.09\\  
		IWUE (kg/m$^3$)& 8.63 & 7.53\\ \bottomrule
	\end{tabular} \label{tbl:relu_trig_dry}
\end{table}

\begin{table}[t]
	\caption{Comparison between the triggered and $\mathbb{P}_{N,M}(d)$ with ReLU neural network for the 2022 season.}
	\small
	\centering
	\begin{tabular}{ccc}
		\toprule
		\textbf{} & \textbf{$\bm{\mathbb{P}_{N,M}(d)}$ with ReLU }& \textbf{Triggered}\\
		\midrule
		Total irrigation (m) & 0.761  & 0.97\\
		IWUE (kg/m$^3$)& 10.65 & 7.78\\   \bottomrule  
	\end{tabular} \label{tbl:relu_trig_wet}
\end{table}

\section{Conclusion}
In conclusion, this study demonstrates the use of ReLU surrogate models to enhance the computational efficiency of mixed-integer Model Predictive Control (MPC)-based irrigation schedulers. By leveraging the mixed-integer linear representation of the ReLU operator, the proposed approach transforms the mixed-integer MPC-based scheduler, which features a quadratic cost function, into a Mixed-Integer Quadratic Program (MIQP). The MIQP represents the simplest class of mixed-integer nonlinear programming  problems, for which global optimization solvers exist.

A case study involving other machine learning surrogate models, specifically the Long Short-Term Memory (LSTM) network, showed that the proposed ReLU-based approach significantly reduces the solution time without compromising the water-saving benefits or the Irrigation Water Use Efficiency (IWUE) enhancement previously established by the mixed-integer MPC-based irrigation scheduler. Additionally, a comparative study with the widely used triggered irrigation scheduling approach demonstrated that the ReLU-based method offers enhanced performance in terms of both total irrigation and IWUE.

These findings confirm that the ReLU-based approach enhances  the computational efficiency while maintaining the effectiveness of mixed-integer MPC-based irrigation schedulers, making it a practical and efficient solution for real-time irrigation management in large-scale agricultural settings.

Despite the promise demonstrated by the proposed approach, several modifications could be made to further enhance its effectiveness. For instance, using estimated hydraulic parameters, derived from an offline parameter estimation approach, as a basis for identifying the ReLU neural network models, plays an important role in reducing parametric uncertainty and effectively addressing plant-model mismatch within scheduler. While this offline strategy provides a good foundation for the real-time implementation of the scheduler, it has limitations in adapting to real-time variations in parameters of the field. To address this limitation, an offset-free MPC approach could be adopted by incorporating a disturbance model. The design of an offset-free MPC, as discussed in~\cite{CAO2021117487}, could serve as a helpful reference in this regard.

\section{Acknowledgements}
Financial support from Natural Sciences and Engineering Research
Council of Canada and Alberta Innovates is gratefully acknowledged.

\appendix

	
	\section{Predicted Yield Calculation}\label{sec:predicted_yield}
	In this study, crop yield is predicted according to the following equation~\cite{bennett2011crop}:
	\begin{equation}
		\label{eq:yield_eqn}
		Y_a =Y_m\left[1- k_y + \left(k_y\times \frac{\text{ET}_{\text{c}}}{\text{ET}_{\text{m}}}\right)\right]
	\end{equation}
	where $Y_a$ is the predicted yield in (kg m$^{-2}$), $Y_m$ is the maximum potential yield in (kg m$^{-2}$),  $\text{ET}_{\text{c}}$ is seasonal crop evapotranspiration (mm), $\text{ET}_{\text{m}}$ is maximum seasonal crop evapotranspiration (mm), and $k_y$ is a crop-specific yield response factor (dimensionless). It is worth noting that this approach assumes that water is the yield-limiting factor. $\text{ET}_{\text{c}}$ is related to $\text{ET}_{\text{m}}$ as follows~\cite{feddes1982simulation}:
	\begin{equation}
		\text{ET}_{\text{c}} = \mathcal{K}(\theta_{\text{v}})\text{ET}_{\text{m}}
	\end{equation}
	where $\mathcal{K}(\cdot)$ is the water stress factor,  which is a function of the volumetric moisture content $\theta_v$. $\mathcal{K}(\cdot)$ is defined as:
	\begin{equation}
		\mathcal{K}(\theta_{\text{v}})=\begin{cases}
			0 & \text{$ \theta_{{\text{v}}_1} \leq \theta_{\text{v}}$}\\
			\frac{\theta_{\text{v}} - \theta_{{\text{v}}_1}}{\theta_{{\text{v}}_2}-\theta_{{\text{v}}_1}} &\text{$ \theta_{{\text{v}}_1} \leq \theta_{\text{v}} \leq \theta_{{\text{v}}_2} $}\\
			1& \text{$ \theta_{{\text{v}}_2} \leq \theta_{\text{v}}\leq\theta_{{\text{v}}_3} $}\\
			\frac{\theta_{\text{v}}-\theta_{{\text{v}}_w}}{\theta_{{\text{v}}_2}-\theta_{{\text{v}}_w}}&\text{$ \theta_{{\text{v}}_w}\leq \theta_{\text{v}}\leq \theta_{{\text{v}}_2} $}
		\end{cases}
	\end{equation}
	where $\theta_{{\text{v}}_1}$ is the volumetric moisture at the anaerobic point, $\theta_{{\text{v}}_2}$ and $\theta_{{\text{v}}_3}$ are the volumetric moisture content values between which optimal water uptake exists, and $\theta_{{\text{v}}_w}$ is  volumetric moisture content at the wilting point $\theta_{\text{wp}}$. $\theta_{{\text{v}}_3} = \bar{\nu}$, $\theta_{{\text{v}}_2} = \underline{\nu}$, and $\theta_{{\text{v}}_w} =\theta_{\text{wp}}$. $\theta_{{\text{v}}_1}$ was calculated as the volumetric moisture content corresponding to a pressure head ($\psi$) of 0.1 m~\cite{capraro2008neural}.
	
	\section{Settings for LSTM network development}\label{sec:lstm_development}
	\subsection{Simulation settings of calibrated 1D equation}\label{sec:settings_RE}
	\begin{table}[H]
		\caption{Simulation settings of calibrated 1D equation}
		\footnotesize
		\centering
		\begin{tabular}{cc}
			\toprule
			\textbf{Depth of soil}& 1.0 m\\
			\hline
			\textbf{Number of spatial nodes}& 31\\
			\hline
			\textbf{Spatial discretization} & \makecell{Top 0.50 m with 21 nodes  \\ Bottom 0.50 m with 11 nodes }\\
			\hline
			\textbf{Time step size} & 30 minutes\\
			\hline
			\textbf{Temporal discretization}& Backward Differentiation Formula\\
			\hline
			\textbf{Process uncertainty}& \makecell{Gaussian with \\Mean = 0.0 \& Standard deviation = 5e-04}\\
			\bottomrule
		\end{tabular} \label{tbl:sim_settings}
	\end{table}
	
	\subsection{Specification of the  ReLU network}\label{sec:lstm_network}  
	\begin{table}[H]
		\caption{Specification of the LSTM network}
		\footnotesize
		\centering
		\begin{tabular}{cc}
			\toprule
			\textbf{Number of hidden layers}& 2\\
			\textbf{Number of neurons}& Layer 1:40, Layer 2: 20\\
			\textbf{Number of epochs}& 40\\
			\textbf{Learning rate} & 0.0001\\
			\textbf{Optimizer}& Adam\\
			\textbf{Time lag}& 1\\
			\textbf{Loss function}& Mean squared error\\
			\bottomrule
		\end{tabular} \label{tbl:hyper_pars}
	\end{table}

	\section{Parameters of formulations $\mathbb{P}_{\text{M}}(y)$}\label{sec:tuning_parameters}
	\begin{table}[H]
		
		\begin{center}
			\caption{Parameters of formulations $\mathbb{P}(y)$ and $\mathbb{P}_{\text{M}}(y)$}
			\footnotesize
			\begin{tabular}{cc}
				\toprule
				\textbf{Parameter}& \textbf{Value} \\     \midrule
				$R_u$ & 9000\\    
				$R_c$ & 1000\\     
				$z_r$ (m) & 0.50 and 1.00 \\   
				{$\bar{Q}$ (MZ$_1$, MZ$_2$, MZ$_3$)} & 22000000\\  
				{$\underline{Q}$ (MZ$_1$, MZ$_2$, MZ$_3$)} & 20000000\\  
				{$\theta^{\text{FC}} / \bar{\nu}$ (MZ$_1$, MZ$_2$, MZ$_3$)} & 0.280, 0.280, 0.300\\
				{$\theta^{\text{WP}}$ (MZ$_1$, MZ$_2$, MZ$_3$)} & 0.120, 0.120, 0.160\\     
				{$\underline{\nu}$ (MZ$_1$, MZ$_2$, MZ$_3$)} for MAD = 50\% & 0.200, 0.200, 0.230\\     \bottomrule
			\end{tabular} \label{tbl:parameter_values}
		\end{center}
	\end{table}
	Note that the specific values of $\theta^{\text{FC}}$ and  $\theta^{\text{WP}}$ corresponding to the soil texture of the MZs were obtained from Reference~\cite{huffman2012irrigation}.
	\section{Uncontrolled inputs for the simulation period}
	
	\subsection{Temperature and reference evapotranspiration}\label{section:weather_data}
	\begin{figure}[H]
		\centerline{\includegraphics[width=0.95\columnwidth]{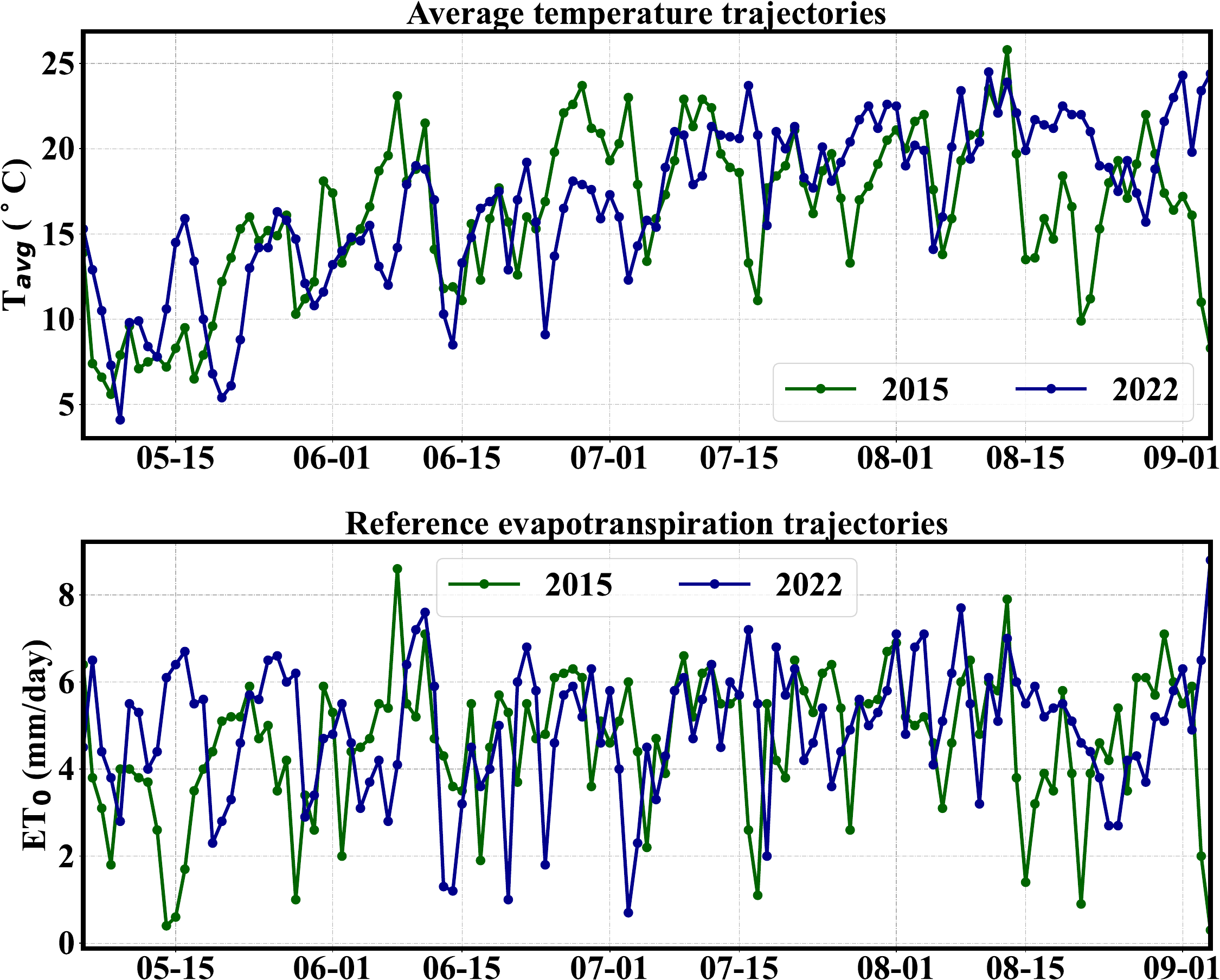}}
		\caption{} 
		\label{fig:weather_data}
	\end{figure}
	
	\subsection{Crop coefficient calculation}\label{sec:crop_coeff}
	The coefficient ($\text{K}_{\text{c}}$)  of soft spring wheat was calculated as follows~\cite{bennett2011crop}:
	\begin{multline}
		\label{eq:kc_relation}
		\text{K}_{\text{c}}(g) = -0.0207 + 0.00266g + \left(4.7\times10^{-8}\right)g^2  -  \left(2.0\times10^{-9}\right)g^3 + \left(2.70\times10^{-13}\right)g^4
	\end{multline}
	where $g$ is the cumulative growing-degree days (GDD). GDD is calculated as follows:
	\begin{equation}
		\text{GDD} = \text{T}_{\text{avg}} - \text{T}_{\text{base}}
	\end{equation}
	where $\text{T}_{\text{avg}}$ is the daily average/mean temperature and $\text{T}_{\text{base}}$ is the base temperature below which crop growth ceases (5\textdegree C). 
	
\section{Extended Kalman Filter Design}\label{sec:ekf_design}
For a particular MZ, the EKF is designed as follows:

\textbf{Initialization}

The EKF is initialized with a guess of the state vector $\hat{x}^{+}_0$, its covariance matrix $P^{+}_0=15.9\mathbb{I}_{21}$,  covariance of the process disturbance $Q=0.05\mathbb{I}_{21}$, and the covariance of the measurement noise $R=19.25$.  

\textbf{Prediction Step}

At time instant $k+1$, $x$ and $P$ are predicted as follows:
\begin{equation}
	\label{eq:prediction}
	\hat{x}^{-}_{k+1}=\mathcal{F}(\hat{x}^{+}_{k},u_k,\bm{\phi})
\end{equation}
\begin{equation}\label{eq:propagation}
	P^{-}_{k+1} = A_k P^{+}_{k} A_k^T + Q
\end{equation}
where $A_k=\frac{\partial \mathcal{F}}{\partial x}\big|_{\hat{x}^{+}_{k},u_k}$ 

\textbf{Update Step} 

Using the soil moisture observation $o_{k+1}$ at time $k+1$, $x$ and $P$ are updated as follows:
\begin{equation}
	\begin{aligned}
		\hat{y}^{-}_{k+1}=\mathcal{H}(\hat{x}^{-}_{k+1},\bm{\phi})
	\end{aligned}
\end{equation}
\begin{equation}
	\begin{aligned}
		G_{k+1} = P^{-}_{k+1} C_{k+1}^T[CP^{-}_{k+1}C_{k+1}^T + R]^{-1}
	\end{aligned}
\end{equation}
\begin{equation}
	\begin{aligned}
		\hat{x}^{+}_{k+1}= &\hat{x}^{-}_{k+1} + G_{k+1}[o_{k+1}-{\mathcal{M}}\hat{y}^{-}_{k+1}]
	\end{aligned}
\end{equation}
\begin{equation}
	P^{+}_{k+1}=[I-G_{k+1}C_{k+1}]P^{-}_{k+1}
\end{equation}
\begin{equation}
	\begin{aligned}
		\hat{y}^{+}_{k+1}=\mathcal{H}(\hat{x}^{+}_{k+1},\bm{\phi})
	\end{aligned}
\end{equation}
where $\mathcal{M}$ serves as a selection matrix utilized to choose the soil moisture contents in $\hat{y}$ that collectively contribute to the soil moisture observation $o$ and $C_{k+1}=\frac{\partial \mathcal{H}}{\partial x}\big|_{\hat{x}^{+}_{k+1}}$.

\end{document}